\documentclass[dvips]{article}
\usepackage{icrctc07}

\title{Geometry reconstruction of fluorescence detectors revisited}
\shorttitle{Geometry reconstruction of fluorescence detectors revisited}
\authors{D.Kuempel$^{1}$, K.-H. Kampert$^{1}$, M. Risse$^{1}$.}
\shortauthors{D. Kuempel and et al}
\afiliations{$^1$Physics Department, University of Wuppertal, Gaussstr. 20, D-42119 Wuppertal, Germany}
\email{kuempel@physik.uni-wuppertal.de}

\abstract{The experimental technique of fluorescence light observation is used in
current and planned air shower experiments that aim at understanding
the origin of ultra-high energy cosmic rays. In the fluorescence technique,
the geometry of the shower is reconstructed based on the correlation
between viewing angle and arrival time of the signals detected by the telescope.
The signals are compared to those expected for different shower geometries
and the best-fit geometry is determined. The calculation of the expected signals
is usually based on a relatively simple function which is motivated by basic
geometrical considerations. This function is based on certain assumptions
on the processes of light emission and propagation through the atmosphere.
For instance, the fluorescence light is assumed to propagate with vacuum speed
of light. We investigate the validity of these assumptions and provide corrections
that can be used in the geometry reconstruction. The impact on reconstruction
parameters is studied. The results are also relevant for hybrid observations where
the shower is registered simultaneously by fluorescence and surface detectors.}

\begin{document}
\maketitle

\section{Introduction}

\begin{figure}[htbp]
\centering
\includegraphics[width=6.5cm]{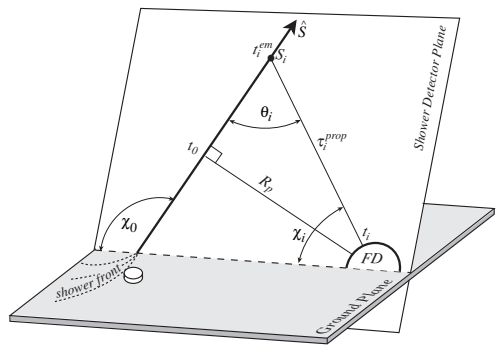}
\caption{Shower/Detector Geometry}
\label{sdp}
\end{figure}

Since the 60's, when the use of extensive air shower (EAS) fluorescence light yield for the ultra high energy cosmic rays (UHECR's)
detection was first proposed, many past, current and future experiments \cite{Abbasi02,Abraham04,Kasahara05} utilize the effect to get a clue about the origin of cosmic rays. \\
First applied for the Utah Fly's Eye detector \cite{Baltrusaitis85} the emitted fluorescence light of EAS is used to reconstruct the shower geometry. The standard single-eye fitting procedure for the shower core location and direction starts with the determination of the plane containing the shower axis and the center of the eye (cf.\ Fig.\ \ref{sdp}). This so called shower-detector plane (SDP) fit uses tube pointing directions, together with signal integrals. To determine the shower orientation within the SDP a correlation between viewing angle $\chi_i$ and firing times of the tubes are used and a best-fit geometry with the expected arrival time at the telescope is accomplished. The calculation of the expected light arrival time is motivated by basic geometrical considerations.

Let $t_i^{em}$ be the time of light emission at point $S_i$ on the shower axis. Then, $t_{i}^{em}$ is calculated as

\begin{eqnarray}
t_{i}^{em}=t_{0}-\frac{R_{p}}{c\tan \theta _{i}}\mbox{ ,}
\label{emission}
\end{eqnarray}

where $t_0$ is the time at which the shower passes the closest point at distance $R_p$ to the detector and $\theta_{i}$ is the angle in the SDP which the shower axis makes with the $i^{th}$ pixel viewing towards $S_i$ (cf.\ Fig.\ \ref{sdp}). The shower front  is assumed here to propagate with the speed of light in vacuum $c$. The time $t_i$ when the light reaches the telescope at pixel $i$ viewing towards $S_i$ is then given by:

\begin{eqnarray}
t_{i} &=& t_{i}^{em}+ \tau_{i}^{prop} \nonumber \\
t_{i} &=& t_{i}^{em}+\frac{R_{p}}{c_{i}^{\prime }\sin \theta _{i}} 
\label{arrive}
\end{eqnarray}

Where $\tau_{i}^{prop}$ is the time of propagation from $S_{i}$ towards the telescope. Assuming also a propagation speed of $c_{i}^{\prime}=c$ for fluorescence light we obtain with (\ref{emission}) and (\ref{arrive}) the ``classical'' formula used so far \cite{Sokolsky89},

\begin{eqnarray}
t_{i} & = & t_{0}+ \frac{R_{p}}{c}\left( \frac{1}{\sin \theta _{i}}-\frac{1
}{\tan \theta _{i}}\right) \nonumber  \\
 &=& t_{0}+\frac{R_{p}}{c}\tan \left( \frac{\chi _{0}-\chi _{i}}{2} \right) \mbox{ ,}
\label{arrival_old}
\end{eqnarray}

where $\chi_0$ and $\chi_i$ are angles within the SDP, see Fig.\ \ref{sdp}. The best fit parameters $R_p$, $t_0$ and $\chi_0$ in Equation (\ref{arrival_old}) are then found by minimizing a $\chi^2$-function. The uncertainty of the three parameters depends on the particular shower geometry and is propagated also for the determination of primary energy and depth of shower maximum. In Eqn.\ (\ref{arrival_old}), it is assumed that everything

\begin{itemize}
\item propagates with vacuum speed of light,
\item takes place instantaneously 
\item propagates on straight lines.
\end{itemize}

We check the validity of these assumptions and provide corrections to Eqn.\ (\ref{arrival_old}).

\section{Reduced speed of light}
Using high-speed electronics with trigger times in the scope of ns, the derivation of the expected light arrival time by basic geometrical considerations has to be revisited. The propagation speed of fluorescence light $v=c/n$ is reduced by an index of refraction $n>1$ which is again a function of the traversed medium and wavelength $\lambda$. The main emission lines of fluorescence light cover a wavelength range between 300 and 400 nm. Within that interval a change of $n(\lambda)$ is $<3 \%$ and negligible. Knowing the local density in air we can calculate the index of refraction $n$ as follows \cite{Heck98}:

\begin{eqnarray}
 n(h) &=& 1 + 0.000283 \cdot \frac{\rho (h)}{\rho(0)} \nonumber \\
 & = & 1 + 0.000283\cdot \frac{c_1}{b_1}\frac{b_j}{c_j} \cdot e^{-h/c_j} 
\label{n_of_h}
\end{eqnarray}

where $h$ is the height a.s.l and $b_j$ as well as $c_j$ atmospheric parameters for the U.S. standard atmosphere given in \cite{Heck98}.

\begin{figure}[t]
\centering
\includegraphics[width=6.5cm]{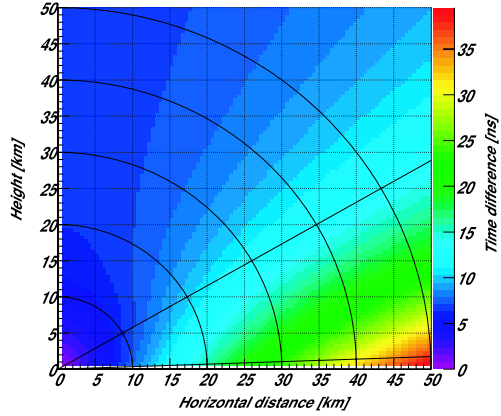}
\caption{Arrival time difference $|t_{real}-t_{vacuum}|$. The telescope is located at position [0,0] at 1416~m  a.s.l.\ (corresponding to the altitude of the Pierre Auger Observatory). The straight lines indicate the field of view between 1 and 31~deg}
\label{matrix}
\end{figure}

To estimate the impact of a realistic atmosphere we calculated the difference of the light arrival times between the cases of vacuum and realistic speed of light from different parts of the atmosphere as shown in Fig.\ \ref{matrix}. We can see that the time difference is larger for light propagating near the earth surface as we would expect and that differences of more than 20-25~ns can occur. Furthermore, we see that coming in showers, with respect to the telescopes, are expected to have a more constant offset contrary to going away showers.

\section{De-excitation lifetimes}

\begin{figure}[t]
\centering
\includegraphics[width=6.5cm]{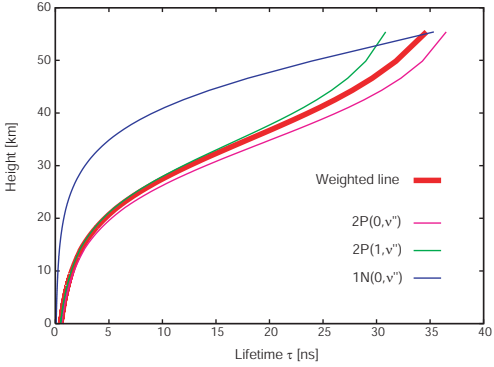}
\caption{Lifetime of individual transitions as a function of height a.s.l.\ for dry air. The thick line indicates the weighted lifetime according to different intensity fractions. The width of that line roughly corresponds to a $\pm$40 K temperature change.}
\label{life_vs_height}
\end{figure}

Another additional delay to the expected arrival time are excitation and de-excitation processes within the shower development induced by low energy electrons and positrons ($\sim 40$ MeV). Almost all of the air fluorescence in the wavelength range between 300 nm and 400 nm originates from transitions of molecular nitrogen $N_2$ or molecular nitrogen ions $N^+_2$ which make $\sim$78\% of the air composition. Excitation times are around 10$^{-15}$~ns and, thus, negligible for our purposes. De-excitation lifetimes can be of the order of 30-40 ns but are affected by quenching \cite{Waldenmaier06}. Here excitated states transfer their energy into rotations, vibrations or translations of other molecules without emitting optical photons. As a consequence one has to introduce an additional radiationless deactivation term $\tau_q$. The inverse mean lifetime $\tau_\nu$ becomes

\begin{equation}
 \frac{1}{\tau_\nu (p,T)} = \frac{1}{\tau_{0_\nu}} + \frac{1}{\tau_{c_\nu}(p,T)},
\end{equation}

where $1/\tau_{0_{\nu}}$ is the sum over all constant transition probabilities. The expected de-excitation lifetimes of the main transitions as a function of altitude are shown in Fig.\ \ref{life_vs_height}.

The functional form of the weighted line in Fig.\ \ref{life_vs_height} can be parameterized in good approximation by

\begin{equation}
 \tau_\nu(h) = \frac{\tau_{0_{\nu}}}{\alpha\cdot e^{-h/H}+1}
\label{tau}
\end{equation}

with $\tau_{0_{\nu}}=37.5$ ns, $H=8005$ m and $\alpha=95$.\\

With Eqn.\ (\ref{emission}), (\ref{arrive}) and (\ref{tau}) the new expected arrival time $t_i$ can then be written in the form

\begin{eqnarray}
t_i  &=&  t_{0}+\frac{R_{p}}{c_{i}^{\prime}}\left(\frac{1}{\sin (\chi _{0}-\chi _{i})} \right) \nonumber\\
& & - \frac{R_{p}}{c} \left( \frac{1}{\tan (\chi _{0}-\chi _{i}) } \right) + \tau_\nu (h) 
\label{arrival_new}
\end{eqnarray}

$c_{i}^{\prime}$ denotes an averaged reduced speed of light for the particular path $i$. Eqn.\ (\ref{arrival_new}) replaces the classical formula Eqn.\ (\ref{arrival_old}) for fitting the shower geometry. Studies on simulated CORSIKA events taking into the Pierre Auger detector MC show, that differences in the arrival time of 20-30 ns and more can occur. The fit parameters $R_p$, $t_0$ and $\chi_0$ are more affected for coming in shower and differences of 5 m, 30 ns and 0.05 deg emerge, respectively.

\section{Bending of light}
In addition to the time delay, the light path also changes according to Fermats principle resulting in an abberation of the viewing angle $\chi_i$. Snell's law states that the ratio of the sines of the angles of incidence $\alpha_{real}$ and refraction $\alpha_{obs}$ is equal to the inverse ratio of the indices of refraction 

\begin{equation}
n_i \sin\alpha_{real} = n_r \sin\alpha_{obs}.
\end{equation}

\begin{figure}[h]
\centering
\includegraphics[width=7cm]{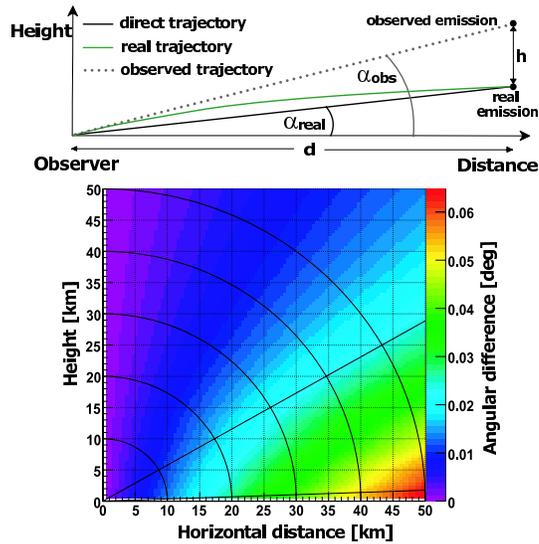}
\caption{\textit{(Top)} Illustration of the path deviation and the resulting impact on the observed emission point. \textit{(Bottom)} Arrival angle difference between direct and curved path for a realistic atmosphere. The different colors indicate the difference $\alpha_{obs} - \alpha_{real}$ in deg.}
\label{diff_angle}
\end{figure}

Consequently the telescope ``detects'' the light higher in the atmosphere than it was actually produced. The angle difference is shown in Fig.\ \ref{diff_angle}.

\begin{figure}[h]
\centering
\includegraphics[width=7cm]{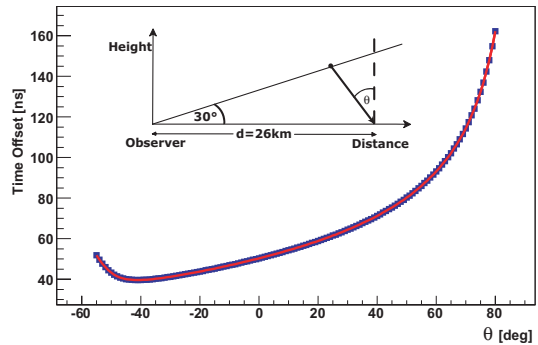}
\caption{Expected time offset as a function of $\theta$ for the toy model. The impact point is always 26 km away with varying starting points along the maximum viewing angle, here 30$^\circ$.}
\label{offset_real}
\end{figure}

An angular difference of $\alpha_{obs} - \alpha_{real}\sim0.05$~deg implies a shifted observed emission point of about $h=30$~m higher at a distance of $d=30$~km. That could cause a delay of $\sim$100~ns for the expected impact time on ground. This is particularly important for so called hybrid observations where the same EAS is detected by the ground array and a fluorescence detector realized at the Pierre Auger Observatory. The resulting relative timing offset has to be taken into account for an accurate reconstruction. In order to estimate the impact of bended and delayed fluorescence light on the time offset a toy model is used. This toy model simulates air showers in vacuum and a realistic atmosphere and determines the expected impact time on ground via the detected fluorescence light at the telescope. Other effects like light attenuation, telescope characteristics etc.\ are neglected. The expected impact time for a realistic atmosphere minus the impact time for a vacuum atmosphere (time offset) as a function of $\theta$ is shown in Fig.\ \ref{offset_real}. There is an increase for inclined showers mainly caused by bended fluorescence light. The rise at low $\theta$ is caused by the absence of quenching effects at high altitudes.

\section{Conclusion}
We studied the impact and influences of realistic light propagation within shower reconstruction from fluorescence light measurements. Differences in the arrival time of 20 - 30 ns and more can occur, which affects the geometry and, to a minor extend, the profile reconstruction. For certain geometries the time offset is expected to exceed 100 ns. A correction for the ``classical'' fit function is provided that can readily be applied to reconstruct fluorescence detector data.

\section{Acknowledgements}
We would like to thank our Auger colleagues, for many fruitful discussions in particular Jose Bellido and Bruce Dawson.

\end{document}